\documentclass{appolb}
\usepackage[utf8]{inputenc}
\usepackage{graphicx}
\usepackage{amsmath}
\usepackage{hyperref}
\usepackage{slashed}
\usepackage{multicol}
\usepackage{multirow}
\usepackage{booktabs}
\usepackage[usenames,dvipsnames]{xcolor}
\usepackage[square,comma,numbers,sort&compress]{natbib}

\newcommand{\eref}[1]{Eq.~(\ref{#1})}

\newcommand{\fref}[1]{Fig.~\ref{#1}}

\definecolor{violet}{RGB}{111,0,255}
\definecolor{dgreen}{rgb}{0.1,0.50,0.1}

\begin{document}

\title{On the quark-gluon vertex at non-vanishing temperature
\thanks{Presented at Excited QCD 2018, 11-15 March 2018, Kopaonik, Serbia}
}
\author{
Romain~Contant$^\dagger$, 
Markus~Q.~Huber$^{\ddagger,\dagger}$, 
Christian~S.~Fischer$^\ddagger$, 
Christian~A.~Welzbacher$^{\ddagger,}\setcounter{footnote}{5}\thanks{Currently at: Deutscher Wetterdienst (DWD), Offenbach, Germany}$, 
Richard~Williams$^\ddagger$
\address{
$^\dagger$Institute of Physics, University of Graz, NAWI Graz, Universit\"atsplatz 5, 8010 Graz, Austria\\
$^\ddagger$Institut f\"ur Theoretische Physik, Justus-Liebig--Universit\"at Giessen, 35392 Giessen, Germany
}
}

\maketitle

\begin{abstract}
We perform a semi-perturbative calculation of the quark-gluon vertex inspired from the three-loop expanded 3PI effective action and investigate the relative strengths of the chirally symmetric/broken tensor structures below and above the crossover.
\end{abstract}


\section{Introduction}\label{sec:introduction}
The quark-gluon vertex is the fundamental link between the matter and the gauge sector of quantum chromodynamics and thus plays a pivotal role in functional studies of the strong interaction.
Approximations using only an effective dressing of the tree-level component have proven useful for studies of bound-states e.g.~\cite{Eichmann:2016yit,Hilger:2017jti}, but the actual structure of the quark-gluon interaction is more complicated as indicated by several recent studies \cite{Fischer:2009jm,Hopfer:2013np,Williams:2014iea,Mitter:2014wpa,Sanchis-Alepuz:2015qra,Williams:2015cvx,Aguilar:2016lbe,Cyrol:2017ewj,Sternbeck:2017ntv,Aguilar:2018epe}.

At non-vanishing temperature, little is known about the details of the quark-gluon interaction. Hence our aim is to explore this quantity in a semi-perturbative calculation~\cite{Fischer:2009jm,Williams:2014iea,Welzbacher:2016rcv} and thus gauge the relative importance of the different components. Such an approximation significantly reduces the complexity of the calculation, since it is a highly technical challenge to account for the full kinematic dependence of the quark-gluon vertex already in vacuum~\cite{Williams:2015cvx}.

\section{Setup}\label{sec:setup}
\begin{figure}[tb]
\begin{center}
\includegraphics[width=0.5\textwidth]{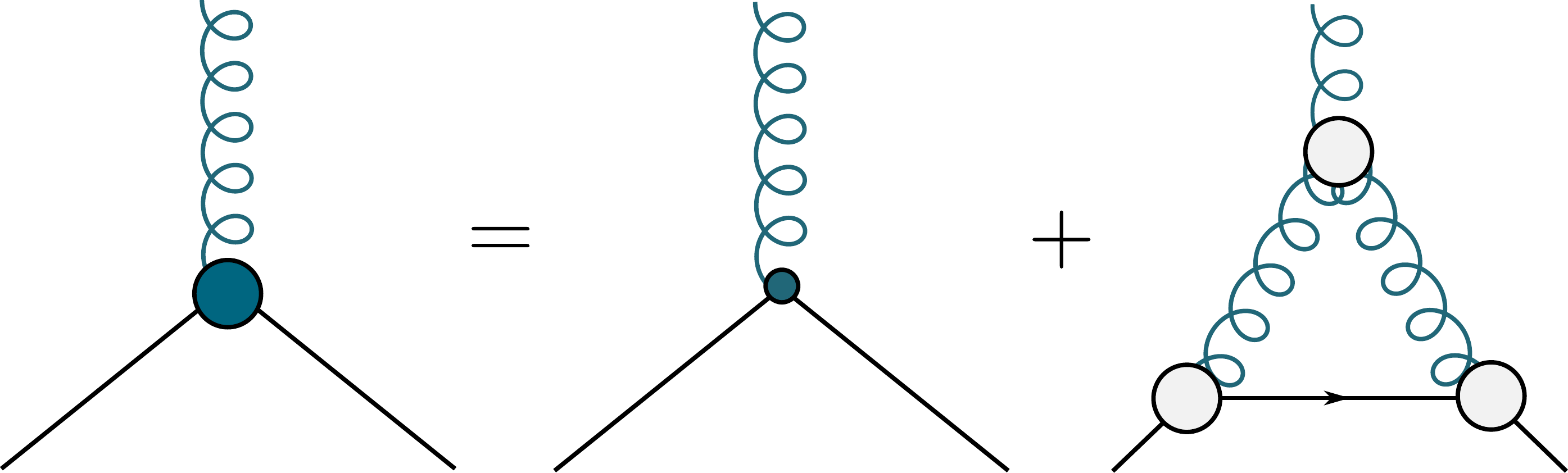}
\caption{The truncated equation of motion for the quark-gluon vertex from the 3PI effective action.
White discs denote the modeled vertices in our semi-perturbative approximation. All internal propagators are dressed with straight lines denoting quarks and wiggly lines denoting gluons.}
\label{fig:qug}
\end{center}
\end{figure}

We perform a semi-perturbative calculation of the quark-gluon vertex inspired from its equation of motion of the three-loop expanded 3PI effective action~\cite{Berges:2004pu,Carrington:2010qq,Williams:2015cvx}.
The so-called Abelian diagram in this equation is suppressed both dynamically and by color factors and is thus neglected herein. The resulting equation is shown in~\fref{fig:qug}.

The vertex equation depends on the quark and gluon propagators and the three-gluon vertex for which we use fixed input.
For the propagators we use results obtained by unquenching lattice results for the gluon propagator \cite{Nickel:2006vf,Fischer:2009wc} by adding the nonperturbative quark-loop \cite{Fischer:2012vc}.
For the quark-gluon vertex in this calculation the following model was used \cite{Fischer:2009wc}:
\begin{align}
\Gamma_{\mu}(k;p,q) &= \gamma_{\mu} \Gamma_\text{mod}(x) \Bigg(\frac{A(p^2) + A(q^2)}{2} \delta_{\mu, i}+\frac{C(p^2) + C(q^2)}{2} \delta_{\mu, 4} \Bigg),\\
\Gamma_\text{mod}(x) &= \frac{d_{1}}{x+d_2} + \frac{x}{\Lambda^2 + x} \left(\frac{\alpha(\mu)\beta_{0}}{4 \pi}\textrm{ln}\left(\frac{x}{\Lambda^2} + 1\right)\right)^{2 \delta}.
 \label{eq:qglvertModel}
\end{align}
Here, $\delta=-9N_c/(44N_c-8N_f)$ is the anomalous dimension of the ghost and $\beta_0=(11N_c-2N_f)/3$ the first coefficient of the beta function.
$\alpha(\mu)$ is given by $0.3$, $d_1$ by $7\,\text{GeV}^2$, $d_2$ by $0.5\,\text{GeV}^2$ and $\Lambda^2$ by $1.96\,\text{GeV}^2$.
Temperature dependence enters via the quark propagator dressing functions $A(p^2)$ and $C(p^2)$. We use the corresponding results for $N_f=2$ from Ref.~\cite{Contant:2017gtz}, where a bare quark mass slightly different to that in~Ref.~\cite{Fischer:2012vc} was employed.

For the three-gluon vertex we take only its tree-level tensor and dress it with a model motivated by the one from Ref.~\cite{Huber:2017txg}:
\begin{align}
D^{AAA}(k_1,k_2,k_3) = -G(3 s_0)^{3} f_d(k_1^2) f_d(k_2^2) f_d(k_3^2) + \frac{3 s_0}{a + 3 s_0} \frac{G(3 s_0)}{Z(3 s_0)},
\end{align}
where $s_0 = (k_1^2 + k_2^2 + k_3^2)/6$, $f_d(x) =1/(1 + x/\Lambda_{3g}^2)$, $a=1\,\text{GeV}^2$ and $\Lambda_\text{3g}^2=0.01\,\text{GeV}^2$.
$G$ and $Z$ are the ghost and gluon dressing functions, respectively.
This ansatz has the correct UV behavior, given by the second term, and features a sign change at low momenta.

Since we will solve the vertex equation semi-perturbatively rather than self-consistently, the internal ansatz for the quark-gluon vertex plays a special role.
We use a variation of \eref{eq:qglvertModel} that effectively captures the infrared contribution in a dressing of the tree-level tensor $\gamma_{\mu}$:
\begin{align}
\hat{\Gamma}_{\mu}(k;p,q) &= \gamma_{\mu} \hat{\Gamma}_\text{mod}(k^2),\\
 \hat{\Gamma}_\text{mod}(x)&= \frac{d_1^2}{\left(x+d_2\right)^2} + \left(\frac{\alpha(\mu)\beta_{0}}{4 \pi}\textrm{ln}\left(\frac{x}{\Lambda^2} + 1\right)\right)^{\delta}.
\end{align}
Here, $\alpha(\mu)$, $\Lambda$ and $d_2$ have the same values as above and $d_1$ is $0.5\,\text{GeV}^2$.
For the UV exponent the anomalous dimension of the vertex is employed, since in contrast to the model used in the propagator Dyson-Schwinger equations we don't need the additional factor to serve as a renormalization group improvement.

\section{Results}
\label{sec:results}

The quark-gluon vertex has 8 transverse tensors at zero temperature.
At non-vanishing temperature they split up into 24 tensors:
\begin{align}
 &\Gamma^{A\bar \psi\psi,a}_{\mu,ij}(l,k)=i\,g\,T^a_{ij}\Gamma_{\mu}(l,k)=i\,g\,T^a_{ij}\sum_{i=1}^{24}h_i(l,k)\tau^i_\mu(l,k),
\end{align}
where $l^{\mu} = \frac{p^{\mu} + q^{\mu} }{2}$ is the relative quark momentum with $p$ ($q$) the incoming (outgoing) quark momentum. The incoming gluon momentum is $k^{\mu} = q^{\mu} - p^{\mu} $, and the direction of the heat bath is taken to be $u^{\mu}$. Explicit tensors are given in Tab.~\ref{tab:tensors}. We use the framework of~\textit{CrasyDSE}~\cite{Huber:2011xc} for the computation.

\begin{table}[htbp]\scriptsize
	\centering
	\begin{tabular}{@{}llllcllll@{}}
		\toprule
		\multicolumn{9}{c}{Vacuum}\\
		\midrule
		\multicolumn{4}{c}{Chirally Broken} & & \multicolumn{4}{c}{Chirally Symmetric} \\
		\cmidrule{1-4}
		\cmidrule{6-9}
		$\gamma_\mu$ & $\slashed{l}l_\mu$ & $\slashed{k}l_\mu$ & 
		$\frac{\gamma_\mu}{2}[\slashed{l},\slashed{k}]$ & & $ i l_\mu$ & $ i \slashed{l}\gamma_\mu$ &  $ i \slashed{k}\gamma_\mu$ & $ \frac{i}{2}l_\mu[\slashed{l},\slashed{k}]$ \\
		\midrule
		\multicolumn{9}{c}{Finite Temperature}\\
		\midrule
		\multicolumn{4}{c}{Chirally Broken} & & \multicolumn{4}{c}{Chirally Symmetric} \\
		\cmidrule{1-4}
		\cmidrule{6-9}
		$\vec{\gamma}_\mu $ & 
		$\vec{l}_\mu \slashed{\vec{l}}$ & 
		$\vec{l}_\mu \slashed{\vec{k}}$ & 
		$\frac{\vec{\gamma}_\mu}{2}[\slashed{\vec{l}},\slashed{\vec{k}}]$ & & 
		$i \vec{l}_{\mu}$ &
		$i u_\mu$ & 
		$i \vec{\gamma}_{\mu}\vec{\slashed{l}}$ & 
		$i \vec{\gamma_{\mu}}\vec{\slashed{k}}$ \\
		$u_{\mu} \slashed{\vec{l}}$ &
		$u_{\mu} \slashed{\vec{k}}$ &
		$u_\mu \slashed{u}$ &
		$\vec{l_\mu} \slashed{u}$ & &
		$\frac{i}{2}\vec{l}_{\mu}[\vec{\slashed{l}}, \vec{\slashed{k}}]$ & 
		$\frac{i}{2}u_{\mu}[\vec{\slashed{l}}, \vec{\slashed{k}}]$ & 
		$i \vec{\gamma}_{\mu}\slashed{u}$ & 
		$i u_{\mu}[\vec{\slashed{l}}, \slashed{u}]$ \\
		$\frac{\vec{\gamma}_\mu}{2}[\vec{\slashed{l}},\slashed{u}]$ &
		$\frac{\vec{\gamma}_\mu}{2}[\vec{\slashed{k}},\slashed{u}]$ &
		$\vec{l}_\mu \slashed{u}\vec{\slashed{l}}\vec{\slashed{k}}$ &
		$u_\mu \slashed{u}\vec{\slashed{l}}\vec{\slashed{k}}$  & &
		$i \vec{l}_{\mu}[\vec{\slashed{l}}, \slashed{u}]$ & 
		$i u_{\mu}[\vec{\slashed{k}}, \slashed{u}]$ & 
		$i \vec{l}_{\mu}[\vec{\slashed{k}}, \slashed{u}]$ & 
		$i\vec{\gamma}_{\mu}\vec{\slashed{l}}\vec{\slashed{k}}\slashed{u}$ \\
		\bottomrule
	\end{tabular} \caption{A choice for the 24 transverse tensor components of the quark-gluon vertex at non-vanishing temperature (the chirally broken group numbers 1 through 12, whilst the chirally symmetric group numbers 13 through 24).
	For reference, the 8 vacuum components are shown at the top.}
\label{tab:tensors}
\end{table}

For the purpose of presentation, we show spatially symmetric points $\vec{p}^2 = \vec{q}^2 = \vec{k}^2 = 2 s_0 $ with temporal components $ (u \cdot k) = 0$ and $(u \cdot p) = (u \cdot q) = \pi T$.

\begin{figure}[tb]
\begin{center}
\includegraphics[width=0.48\textwidth]{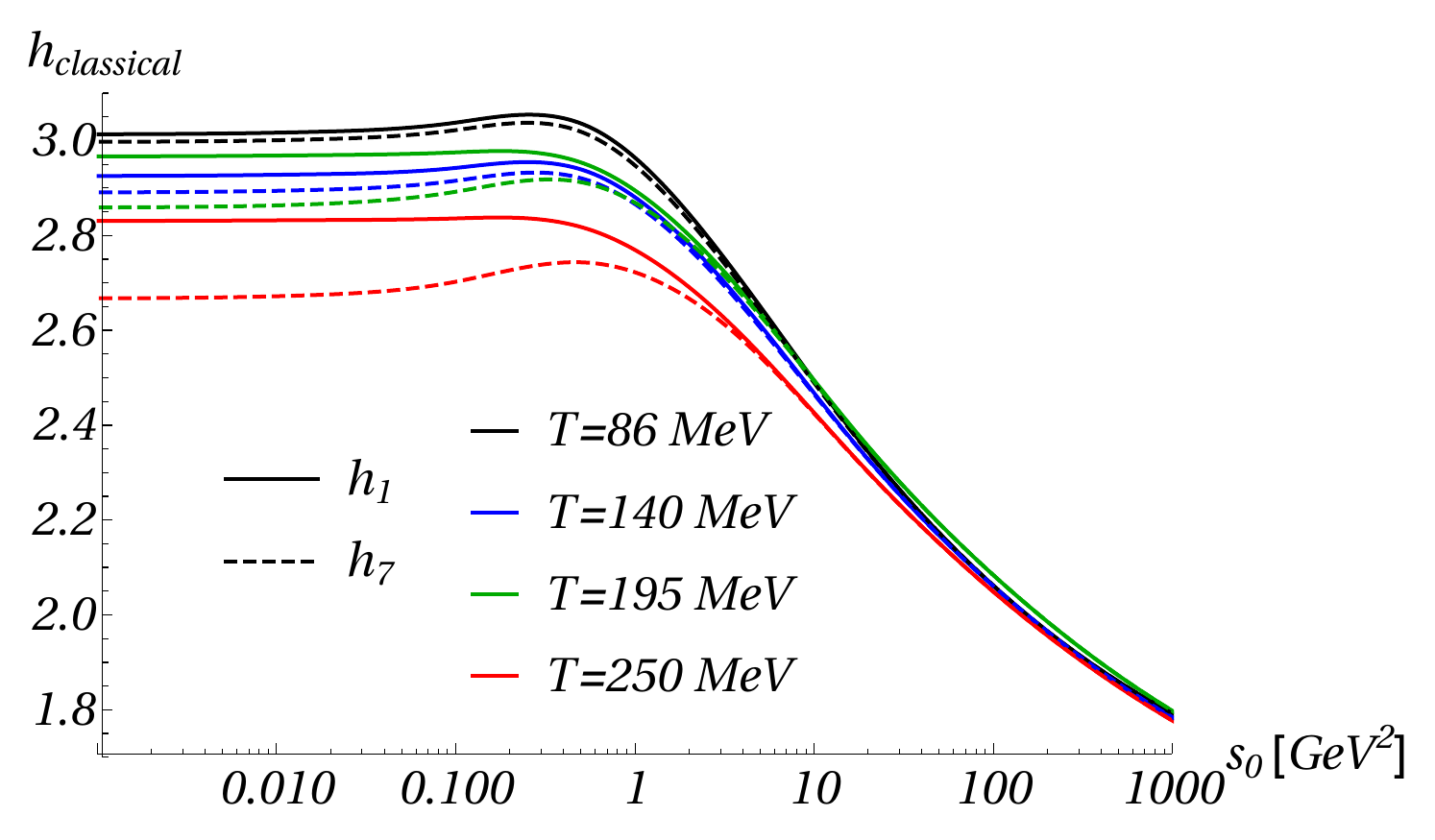}
\hfill
\includegraphics[width=0.48\textwidth]{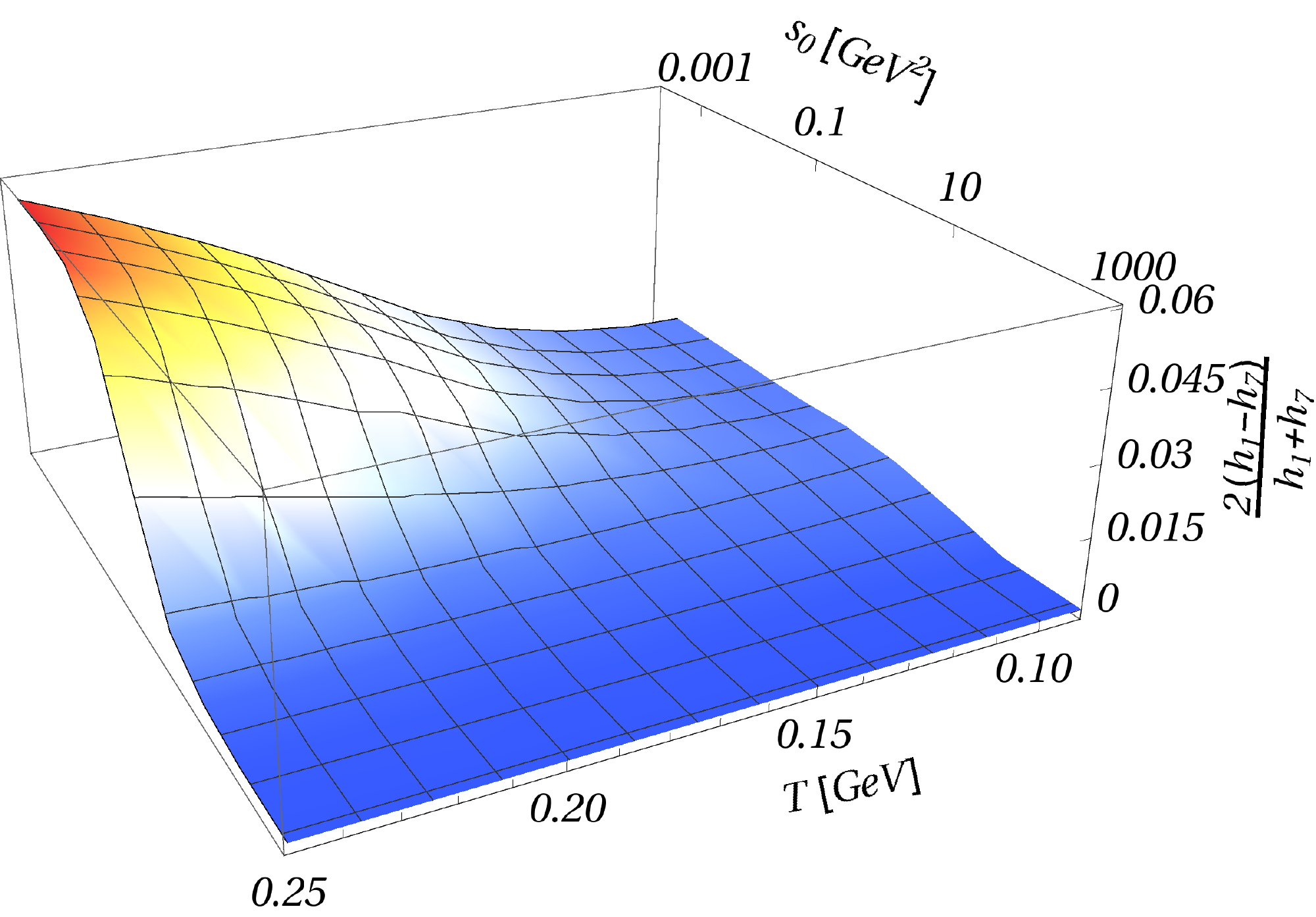}
\caption{Temperature dependence of the classical dressing functions $h_1$ and $h_7$ (left).
In the vacuum they merge to a single dressing function as shown in the plot of the difference between the two dressings (right).}
\label{fig:classical}
\end{center}
\end{figure}

\begin{figure}[tb]
\begin{center}
\includegraphics[width=0.48\textwidth]{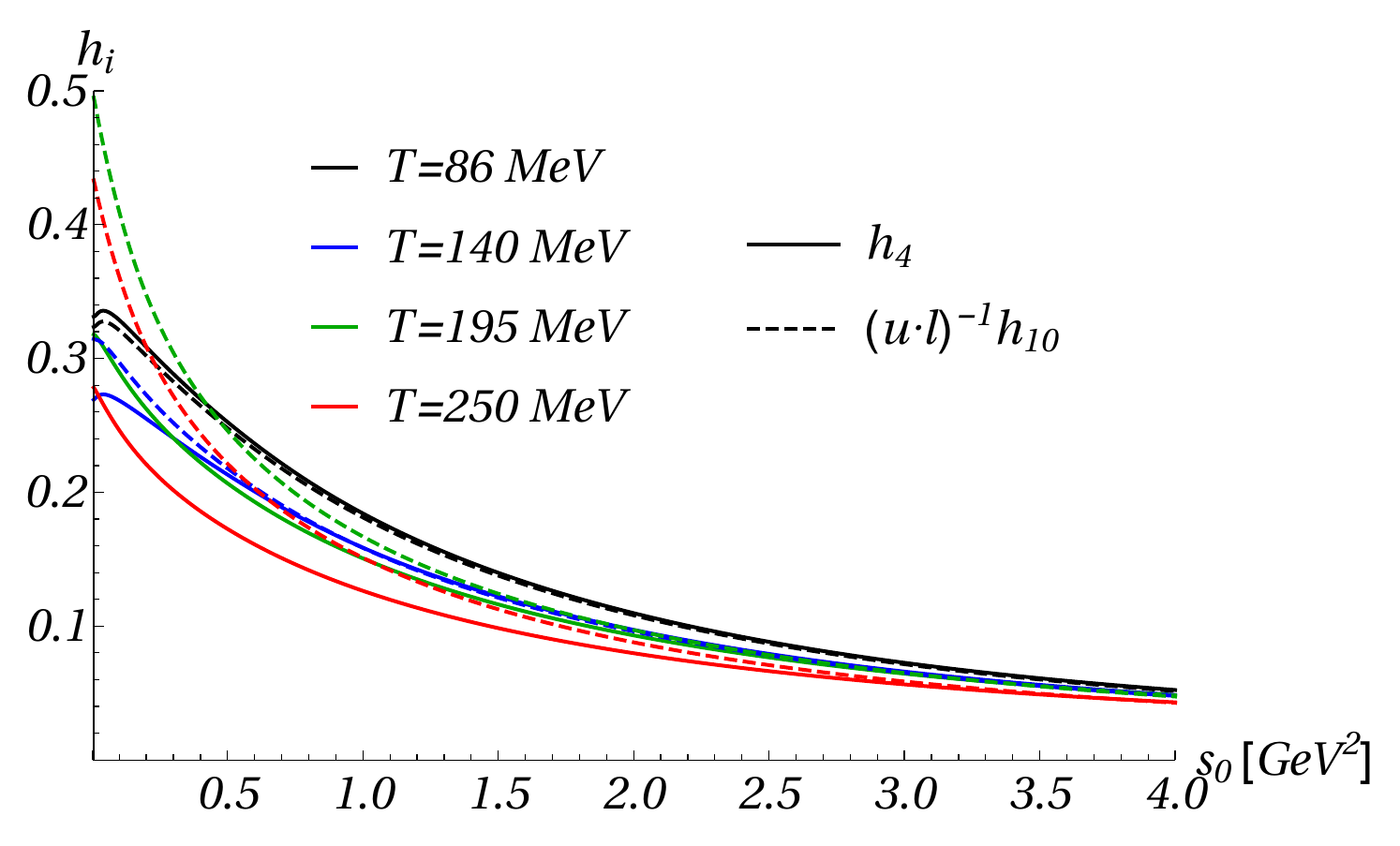}
\hfill
\includegraphics[width=0.48\textwidth]{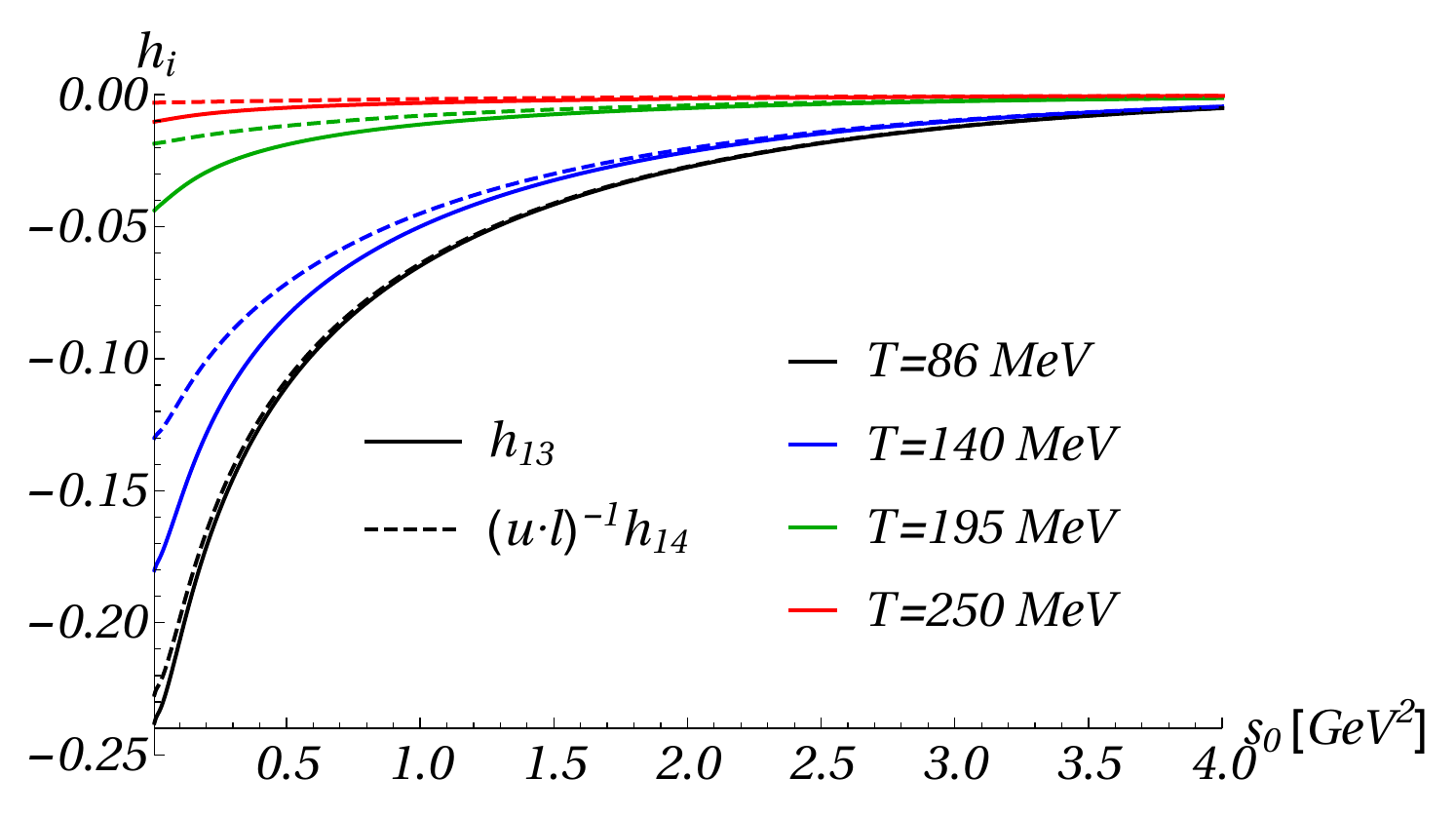}
\caption{Temperature dependence of the dressings $h_4$ and $(u \cdot l)^{-1}h_{10}$ (left) as well as of $h_{13}$ and $(u \cdot l)^{-1} h_{14}$ (right).
The respective pairs join in the vacuum to a single dressing function.}
\label{fig:comparison}
\end{center}
\end{figure}

\begin{figure}[tb]
\begin{center}
\includegraphics[width=0.41\textwidth]{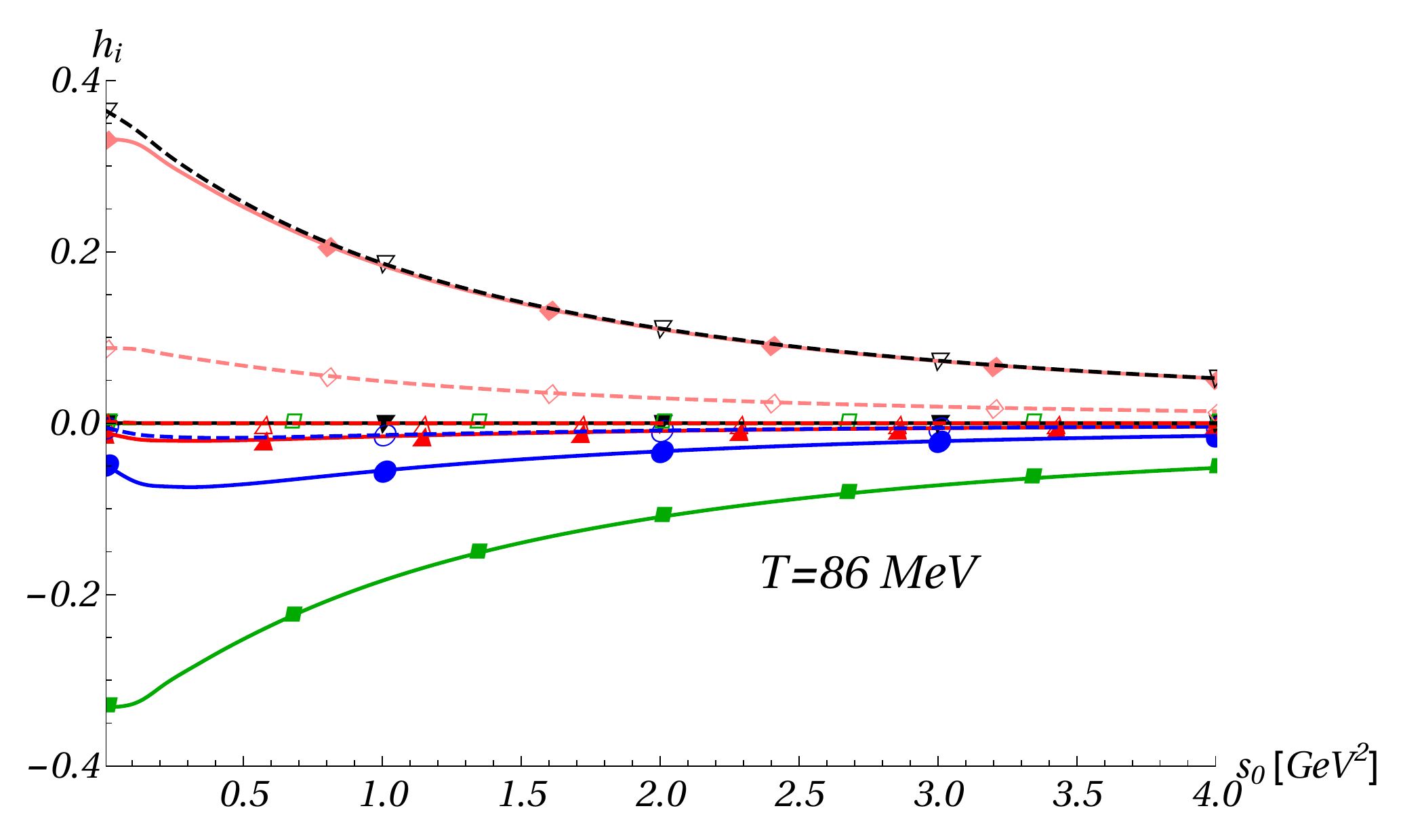}
\includegraphics[trim={0.82cm 1cm 0.8cm 0},clip,width=0.15\textwidth]{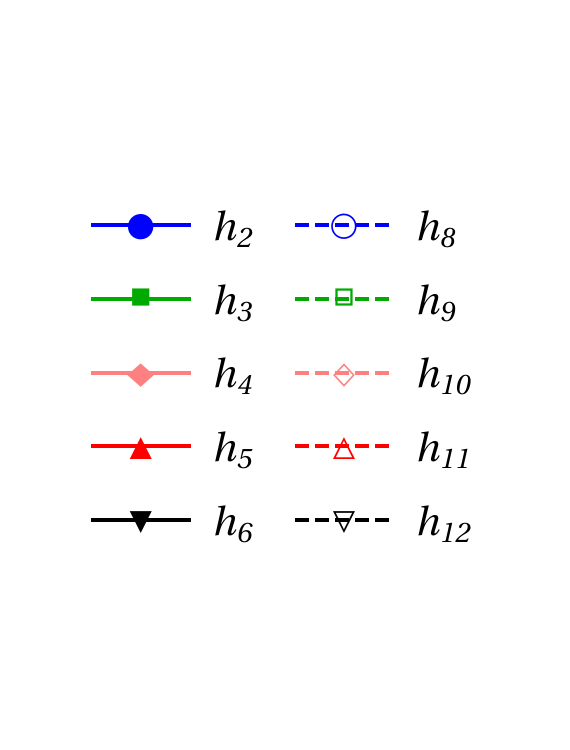}
\includegraphics[width=0.41\textwidth]{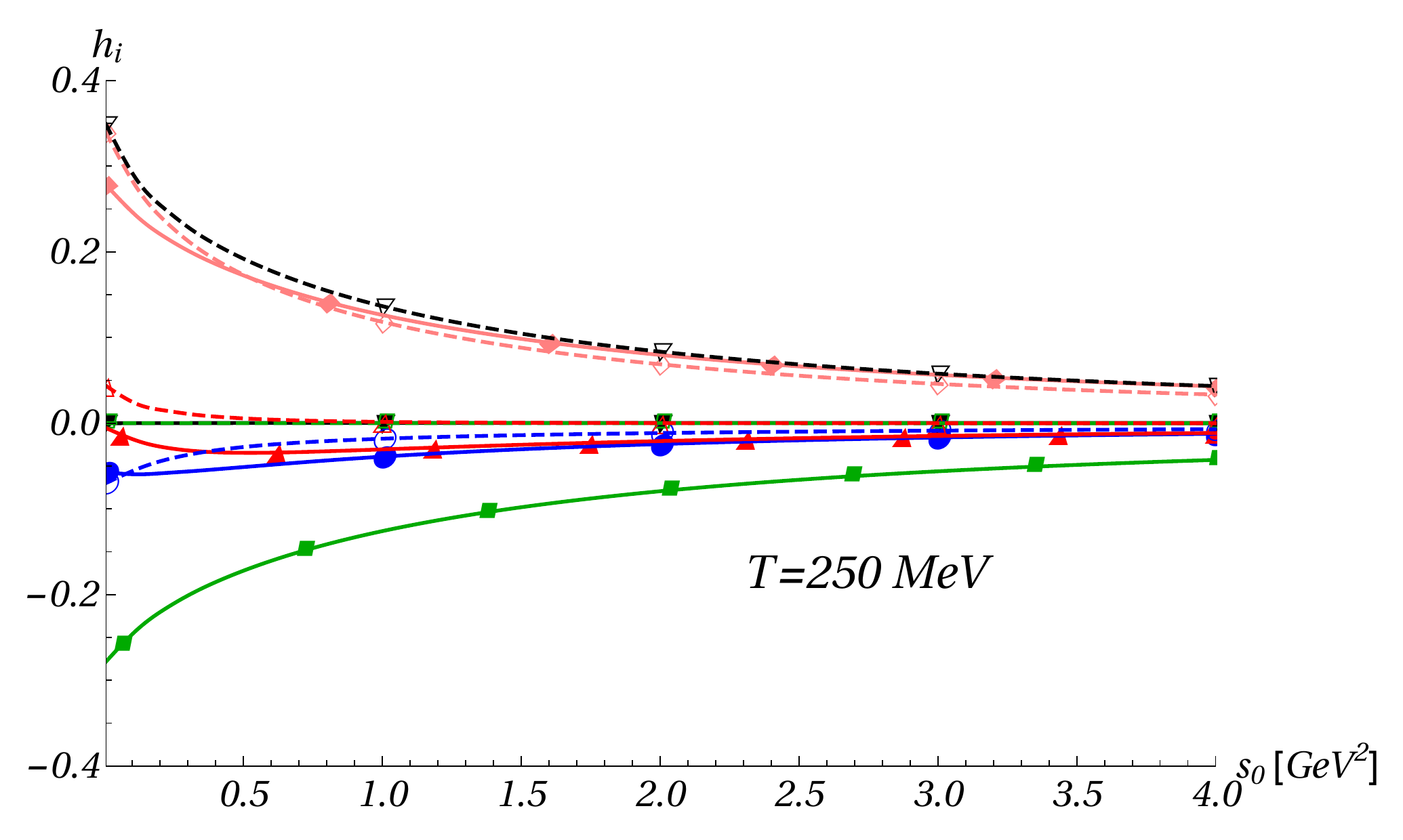}
\includegraphics[width=0.41\textwidth]{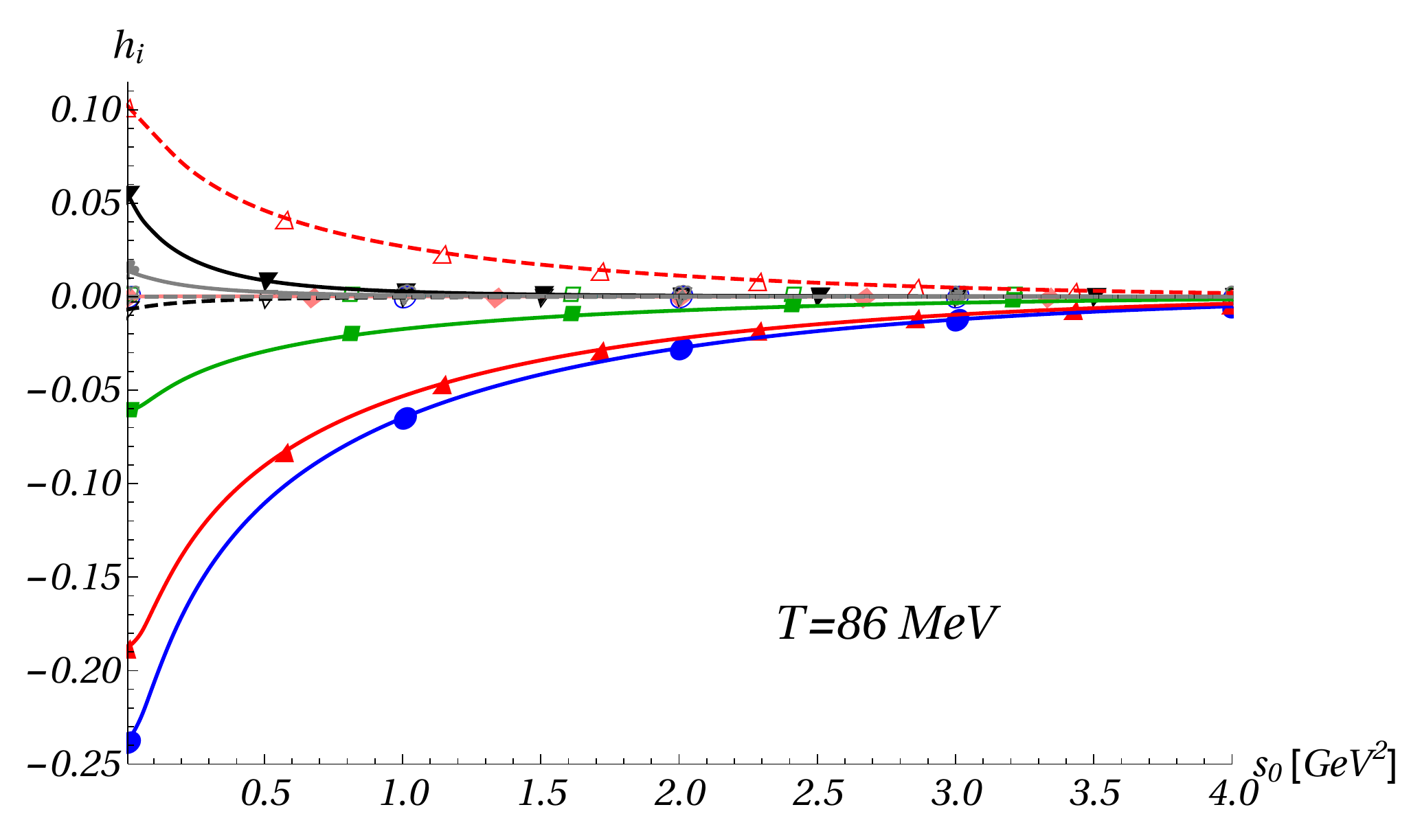}
\includegraphics[trim={2cm 0.5cm 1.72cm 0.5cm},clip,width=0.15\textwidth]{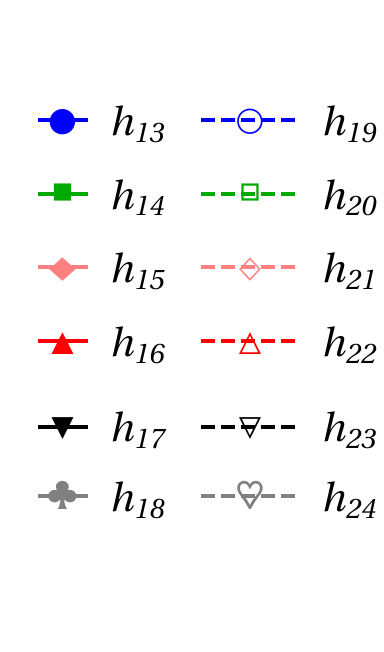}
\includegraphics[width=0.41\textwidth]{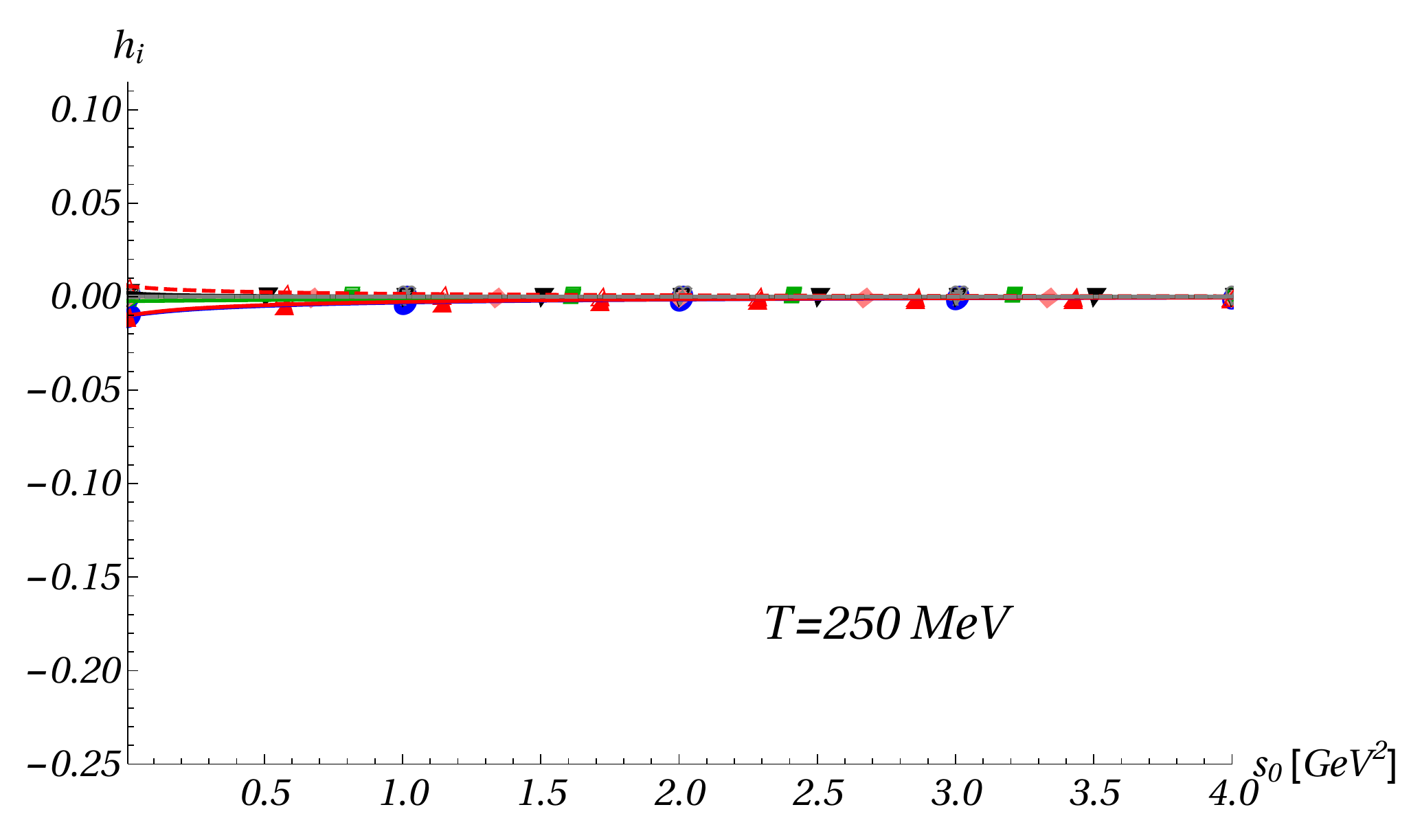}
\caption{All non-classical chirally symmetric (top) and broken (bottom) dressing functions below (left) and above (right) the crossover.}
\label{fig:temperature}
\end{center}
\end{figure}

The temperature dependence of the tree-level dressings is shown in Fig.~\ref{fig:classical} and that of selected dressings in Fig.~\ref{fig:comparison}.
At low temperatures, the dressing functions that stem from a single vacuum tensor become degenerate.
In Fig.~\ref{fig:temperature} one can see that beyond the crossover, which for this $N_f=2$ calculation occurs around $180\,\text{MeV}$, the dressings forbidden by chiral symmetry are suppressed as expected.

\section{Summary}\label{sec:summary}
We presented a semi-perturbative calculation of the quark-gluon vertex at finite temperature, demonstrating the enhancement and suppression of chirally forbidden dressings below and above crossover. Further we see that in the limit of vanishing temperature, the multitude of dressing functions split by the introduction of the heat bath degenerates, thus recovering the vacuum structure.

\section*{Acknowledgments}
Results have been obtained using the HPC clusters at the University of Graz. Funding by the FWF (Austrian science fund) under Contract No.~P 27380-N27, the German Federal Ministry of Education and Research (BMBF) under Contract No.~05P15RGFCA, and the Helmholtz International Center for FAIR within the LOEWE program of the State of Hesse are gratefully acknowledged. 

\bibliographystyle{utphys_mod}
\bibliography{literature_eQCD2018_CONTANT}

\end{document}